\documentclass[11pt,a4paper]{article}

\usepackage{amssymb}
\usepackage{graphicx}
\usepackage{bm}

\usepackage{mathrsfs}
\usepackage{cite}

\begin{document}

\title{The $\beta$-function of supersymmetric theories from vacuum supergraphs: a three-loop example}

\author{S.S.Aleshin\,${}^{a}$, I.O.Goriachuk\,${}^{b}$, D.S.Kolupaev\,${}^b$, K.V.Stepanyantz\,${}^{bc}$ $\vphantom{\Big(}$
\medskip\\
${}^a${\small{\em Institute for Information Transmission Problems RAS,}}\\
\medskip
{\small{\em 127051, Moscow, Russia,}}\\
${}^b${\small{\em Moscow State University, Faculty of Physics,}}\\
{\small{\em Department of Theoretical Physics,}}\\
\medskip
{\small{\em 119991, Moscow, Russia,}}\\
${}^c${\small{\em Moscow State University, Faculty of Physics,}}\\
{\small{\em Department of Quantum Theory and High Energy Physics,}}\\
{\small{\em 119991, Moscow, Russia}}
\medskip
}

\maketitle

\begin{abstract}
We verify a method which allows to obtain the $\beta$-function of supersymmetric theories regularized by higher covariant derivatives by calculating only specially modified vacuum supergraphs. With the help of this method for a general renormalizable ${\cal N}=1$ supersymmetric gauge theory a part of the three-loop $\beta$-function depending on the Yukawa couplings is constructed in the general $\xi$-gauge. The result is written in the form of an integral of double total derivatives with respect to the loop momenta. It is demonstrated that all gauge dependent terms cancel each other in agreement with the general statements. Taking into account that the result in the Feynman gauge (found earlier) coincides with the one obtained by the standard technique, this fact confirms the correctness of the considered method by a highly nontrivial multiloop calculation.
\end{abstract}

\section{Introduction}
\hspace*{\parindent}

Precise calculations of quantum corrections in ${\cal N}=1$ supersymmetric theories are very important for both theory and elementary particle physics \cite{Mihaila:2013wma}. For instance, the unification of running coupling constants in supersymmetric extensions of the Standard Model is the main indirect evidence of supersymmetry existence \cite{Ellis:1990wk,Amaldi:1991cn,Langacker:1991an}. The renormalization group behaviour of these constants is encoded in the $\beta$-functions. Numerous multiloop calculations of the $\beta$-function for ${\cal N}=1$ supersymmetric theories (up to the four-loop approximation) have been made in the $\overline{\mbox{DR}}$-scheme, see., e.g., \cite{Avdeev:1981ew,Jack:1996vg,Jack:1996cn,Jack:2004ch,Jack:2005id,Harlander:2006xq}. Also it is known that there are some schemes (which do not include $\overline{\mbox{DR}}$) in which the $\beta$-function of these theories is related to the anomalous dimension of the matter superfield in all loops by the so-called NSVZ equation \cite{Novikov:1983uc,Jones:1983ip,Novikov:1985rd,Shifman:1986zi}. Such equations can in particular be written for theories with multiple gauge couplings  \cite{Shifman:1996iy,Ghilencea:1999cy}. For some phenomenologically interesting models the NSVZ equations can be found in, e.g., \cite{Shifman:1996iy,Korneev:2021zdz}.

The perturbative derivation of the NSVZ $\beta$-function and of an all-loop prescription for constructing some NSVZ schemes \cite{Stepanyantz:2016gtk,Stepanyantz:2019ihw,Stepanyantz:2020uke} (see also \cite{Stepanyantz:2019lfm}) is based on the fact that the integrals defining the $\beta$-function in supersymmetric theories are integrals of double total derivatives in the momentum space if a regularization is made by the higher covariant derivative method \cite{Slavnov:1971aw,Slavnov:1972sq,Slavnov:1977zf} in the superspace formulation \cite{Krivoshchekov:1978xg,West:1985jx}.\footnote{In the case of using the regularization by dimensional reduction the structure of loop integrals is quite different \cite{Aleshin:2015qqc,Aleshin:2016rrr}.} This feature has first been noted in calculating the lowest quantum correction for ${\cal N}=1$ SQED regularized by higher derivatives in \cite{Soloshenko:2003nc} (the factorization into total derivatives) and \cite{Smilga:2004zr} (the factorization into double total derivatives). Subsequently the factorization into (double) total derivatives was demonstrated in other calculations made for various theories, see, e.g., \cite{Pimenov:2009hv,Stepanyantz:2011bz,Stepanyantz:2011cpt,Buchbinder:2014wra,Buchbinder:2015eva,Shakhmanov:2017soc,Kazantsev:2018nbl}. The all-loop proof of this fact has been done in \cite{Stepanyantz:2011jy,Stepanyantz:2014ima} for the Abelian case and in \cite{Stepanyantz:2019ihw} for the non-Abelian theories. The derivation made in the latter paper allowed to construct a simple method for calculating the $\beta$-function for ${\cal N}=1$ supersymmetric theories regularized by higher covariant derivatives. (Some improvements were also introduced in \cite{Kuzmichev:2019ywn,Stepanyantz:2019lyo}.) Usually such calculations in higher orders are extremely complicated due to a very large number of Feynman (super)diagrams with two external legs of the background gauge (super)field. The new method allows calculating only vacuum supergraphs modified in a special way. More exactly, a special algorithm allows to construct a contribution to the $\beta$-function coming from all superdiagrams  which are obtained from a considered vacuum supergraph by attaching two external gauge legs in all possible ways.  Certainly, a number of such vacuum supergraphs is much less than a number of the two-point ones.

Some explicit calculations with the help of the new method in the two- and three-loop approximations have been done in \cite{Stepanyantz:2019ihw,Kuzmichev:2019ywn,Stepanyantz:2019lyo,Aleshin:2020gec}. Moreover, the corresponding technique has essentially been used for deriving the NSVZ $\beta$-function in \cite{Stepanyantz:2020uke}. Nevertheless, it is so unusual that it is desirable to make some more nontrivial tests. For example, it is possible to obtain gauge dependent contributions to the $\beta$-function and verify their cancellation, which should occur due to some general theorems, see \cite{Batalin:2019wkb} and references therein. Note that the calculations in the nonminimal gauges are as a rule much more complicated than the ones in the minimal gauges. Therefore, a check of the $\beta$-function gauge independence can be considered as a very nontrivial test of the method for making the calculation. Moreover, calculations in nonminimal gauges sometimes reveal some interesting features of quantum corrections. For instance, in the general $\xi$-gauge (unlike the Feynman gauge) the nonlinear renormalization of the quantum gauge superfield becomes essential for calculating the two-loop anomalous dimension of the Faddeev--Popov ghosts \cite{Kazantsev:2018kjx}.

In this paper in the (nonminimal) $\xi$-gauge for a general renormalizable ${\cal N}=1$ supersymmetric gauge theory we calculate a three-loop contribution to the $\beta$-function containing the Yukawa couplings. The corresponding calculation in the (minimal) Feynman gauge has been done in \cite{Kazantsev:2018nbl} by the standard method, and in \cite{Stepanyantz:2019ihw} with the help of vacuum supergraphs. Here we make a similar calculation in the general $\xi$-gauge and verify that the gauge dependence of the corresponding contribution to the $\beta$-function vanishes.

The paper is organized as follows. In Sect. \ref{Section_Method} we briefly recall how one can calculate the $\beta$-function for ${\cal N}=1$ supersymmetric theories regularized by higher covariant derivatives with the help of vacuum supergraphs. Using this method in Sect. \ref{Section_Calculation} the cancellation of the gauge dependent terms in the part of the three-loop $\beta$-function depending on the Yukawa couplings is demonstrated. The results for contributions of separate supergraphs (which appear to be gauge dependent) are presented in Appendix \ref{Appendix_Supergraph_Results}. A brief discussion is made in Conclusion.

\section{Obtaining the $\beta$-function by calculating vacuum supergraphs}
\hspace*{\parindent}\label{Section_Method}

According to \cite{Stepanyantz:2019ihw}, for ${\cal N}=1$ supersymmetric gauge theories regularized by higher covariant derivatives the $\beta$-function defined in terms of the bare couplings can be obtained by calculating specially modified vacuum supergraphs. In the massless limit these (in general, non-Abelian) theories are described by the action

\begin{equation}
S = \frac{1}{2e_0^2}\mbox{Re}\, \mbox{tr} \int d^4x\, d^2\theta\, W^a W_a + \frac{1}{4} \int d^4x\, d^4\theta\, \phi^{*i} (e^{2V})_i{}^j \phi_j + \Big(\frac{1}{6}\int d^4x\, d^2\theta\, \lambda_0^{ijk} \phi_i \phi_j \phi_k + \mbox{c.c.}\Big)
\end{equation}

\noindent
written in terms of ${\cal N}=1$ superfields. This formulation provides manifest ${\cal N}=1$ supersymmetry which remains even at the quantum level. The gauge invariance imposes the constraint

\begin{equation}
\lambda_0^{ijm} (T^A)_m{}^k + \lambda_0^{imk} (T^A)_m{}^j + \lambda_0^{mjk} (T^A)_m{}^i = 0
\end{equation}

\noindent
to the bare Yukawa couplings, where $T^A$ are the generators of the gauge group $G$ in the representation $R$ to which the matter superfields belong. At the quantum level the manifestly gauge invariant effective action can be constructed with the help of the background (super)field method \cite{DeWitt:1965jb,Abbott:1980hw,Abbott:1981ke} which in the supersymmetric case \cite{Grisaru:1982zh,Gates:1983nr} is introduced by the replacement $e^{2V} \to e^{2{\cal F}(V)} e^{2\bm{V}}$. Here $\bm{V}$ is the Hermitian background gauge superfield, and ${\cal F}(V)$ is a special function of the quantum gauge superfield. The form of this function is determined by the nonlinear renormalization of this superfield \cite{Piguet:1981fb,Piguet:1981hh,Tyutin:1983rg}. In the lowest nontrivial approximation the explicit expression for it can be found in \cite{Juer:1982fb,Juer:1982mp}. Even in the two-loop approximation the nonlinear terms in the function ${\cal F}(V)$ are highly important for the renormalization group equations to be satisfied \cite{Kazantsev:2018kjx}.

For introducing the higher covariant derivative regularization we modify the action by adding the higher derivative term $S_\Lambda$, so that

\begin{eqnarray}\label{Regularized_Action}
&&\hspace*{-5mm} S \to S_{\mbox{\scriptsize reg}} = S + S_\Lambda = \frac{1}{2 e_0^2}\mbox{Re}\, \mbox{tr} \int d^4x\,d^2\theta\, W^a \Big[e^{-2\bm{V}} e^{-2{\cal F}(V)}\,  R\Big(-\frac{\bar\nabla^2 \nabla^2}{16\Lambda^2}\Big)\, e^{2{\cal F}(V)}e^{2\bm{V}}\Big]_{Adj} W_a \nonumber\\
&&\hspace*{-5mm} + \frac{1}{4} \int d^4x\,d^4\theta\, \phi^{*i} \Big[\, F\Big(-\frac{\bar\nabla^2 \nabla^2}{16\Lambda^2}\Big) e^{2{\cal F}(V)}e^{2\bm{V}}\Big]_i{}^j \phi_j
+ \Big(\frac{1}{6} \lambda_0^{ijk} \int d^4x\, d^2\theta\, \phi_i \phi_j \phi_k + \mbox{c.c.} \Big).
\end{eqnarray}

\noindent
In this expression the functions $R(x)$ and $F(x)$ are given by the sums of 1 corresponding to the original action and contributions of higher derivative terms. At infinity these functions should rapidly increase, while at $x=0$ they are equal to 1. One more regulator function $K(x)$ with the same properties is also present in the gauge fixing term

\begin{equation}\label{Gauge_Fixing_Action}
S_{\mbox{\scriptsize gf}} =  - \frac{1}{16\xi_0 e_0^2}\, \mbox{tr} \int d^4x\,d^4\theta\,  (\bm{\nabla}^2)_{Adj} V  K\Big(-\frac{\bm{\bar\nabla}^2 \bm{\nabla}^2}{16\Lambda^2}\Big)_{Adj} (\bm{\bar\nabla}^2)_{Adj} V.
\end{equation}

\noindent
In the above equations $\Lambda$ is a parameter with the dimension of mass, which plays the role of an ultraviolet cutoff. In our conventions the covariant derivatives are given by the expressions

\begin{equation}
\nabla_a = \bm{\nabla}_a \equiv D_a;\qquad \bar\nabla_{\dot a} \equiv e^{2{\cal F}(V)} e^{2\bm{V}} \bar D_{\dot a} e^{-2\bm{V}} e^{-2{\cal F}(V)}; \qquad \bm{\bar\nabla}_{\dot a} \equiv e^{2\bm{V}} \bar D_{\dot a} e^{-2\bm{V}}.
\end{equation}

The generating functional should also contain the Faddeev--Popov and Nielson--Kallosh ghosts and the Pauli--Villars determinants needed for removing one-loop divergences which survive after the replacement (\ref{Regularized_Action}),

\begin{eqnarray}\label{Generating_Functional}
&& Z = \int D\mu\; \mbox{Det}(PV, M_{\varphi})^{-1} \big(\mbox{Det}(PV, M)\big)^{T(R)/T(R_{PV})}\,\nonumber\\
&& \qquad\qquad\qquad\qquad\qquad\qquad \times \exp\Big\{i\Big(S_{\mbox{\scriptsize reg}} + S_{\mbox{\scriptsize gf}} + S_{\mbox{\scriptsize FP}} + S_{\mbox{\scriptsize NK}} + S_{\mbox{\scriptsize sources}}\Big)\Big\},\qquad
\end{eqnarray}

\noindent
see \cite{Aleshin:2016yvj,Kazantsev:2017fdc} for details. The constant $T(R)$ is defined by the equation $\mbox{tr}(T^A T^B) = T(R)\delta^{AB}$ under the assumption that for the fundamental representation $T(\mbox{fund.})=1/2$, and $R_{PV}$ is the representation for the Pauli--Villars superfields with the mass $M$.

In this paper we will calculate the gauge $\beta$-function defined in terms of the bare couplings by the equation

\begin{equation}\label{Beta_Bare}
\beta(\alpha_0,\lambda_0,Y_0) \equiv \frac{d\alpha_0}{d\ln\Lambda}\bigg|_{\alpha,\lambda,Y=\mbox{\scriptsize const}},
\end{equation}

\noindent
where $\alpha$ and $\lambda$ are the renormalized gauge and Yukawa couplings, respectively, and $Y$ denotes the whole set of (renormalized) parameters present inside the function ${\cal F}(V)$ together with the renormalized gauge parameter $\xi$. According to \cite{Kataev:2013eta}, the function (\ref{Beta_Bare}) should be distinguished from the $\beta$-function

\begin{equation}\label{Beta_Renormalized}
\widetilde\beta(\alpha,\lambda,Y) \equiv \frac{d\alpha}{d\ln\mu}\bigg|_{\alpha_0,\lambda_0,Y_0=\mbox{\scriptsize const}}
\end{equation}

\noindent
standardly defined in terms of the renormalized couplings. The function (\ref{Beta_Bare}) depends on a regularization but is independent on a renormalization prescription provided the regularization is fixed. The function (\ref{Beta_Renormalized}) depends on both regularization and renormalization prescription and can easily be obtained from $\beta(\alpha_0,\lambda_0,Y_0)$, see, e.g., \cite{Kazantsev:2020kfl} for the three-loop example.

Standardly, to obtain the $\beta$-function one needs to calculate (super)diagrams with two external legs of the background gauge (super)field. In higher orders of the perturbation theory their number become huge. However, the analysis made in \cite{Stepanyantz:2019ihw} revealed that the $\beta$-function (\ref{Beta_Bare}) for ${\cal N}=1$ supersymmetric gauge theories regularized by higher covariant derivatives can be found much easier, by calculating only specially modified vacuum supergraphs. For this purpose it is necessary to follow the algorithm presented below \cite{Kuzmichev:2019ywn,Stepanyantz:2019lyo}:

1. As a starting point we consider a vacuum supergraph with $L$ loops and construct the expression for it using the superfield Feynman rules.

2. To an arbitrary point we insert the factor $\theta^4 (v^A)^2$ and reduce the resulting expression to a momentum integral with the help of standard rules for calculating supergraphs. Here $v^A$ are functions of the space-time coordinates slowly decreasing at a large scale $R\to \infty$. Therefore, all terms containing the derivatives of $v^A$ should be omitted when calculating.

3. In the considered supergraph we mark $L$ propagators with independent (Euclidean) momenta $Q^\mu_i$ and extract the corresponding product of $\delta$-symbols $\prod_{i=1}^L \delta_{a_i}^{b_i}$.

4. In the integrand of the momentum integral we formally replace the product of $\delta$-symbols corresponding to the marked propagators by the expression

\begin{equation}
\sum\limits_{m,n=1}^L \prod\limits_{i\ne m,n} \delta_{a_i}^{b_i}\, (T^A)_{a_m}{}^{b_m} (T^A)_{a_n}{}^{b_n} \frac{\partial^2}{\partial Q_m^\mu \partial Q_{\mu,n}}.
\end{equation}

5. After applying the operator

\begin{equation}
- \frac{2\pi}{r{\cal V}_4} \frac{d}{d\ln\Lambda},\qquad \mbox{where}\qquad {\cal V}_4 \equiv \int d^4x\, (v^A)^2,
\end{equation}

\noindent
to the resulting expression we obtain the contribution to the function $\beta(\alpha_0,\lambda_0,Y_0)/\alpha_0^2$ coming from all superdiagrams which are obtained from the original supergraph by attaching two external lines of the background gauge superfield $\bm{V}$ in all possible ways.

Evidently, by construction, the result is obtained in the form of an integral of double total derivatives in the momentum space.

\section{The gauge independence of the three-loop contribution to the $\beta$-function containing the Yukawa couplings}
\hspace*{\parindent}\label{Section_Calculation}

The calculation of the three-loop $\beta$-function for the general ${\cal N}=1$ supersymmetric gauge theory regularized by higher covariant derivatives is rather complicated and has not yet been done directly. However, some its parts have already been made. In particular, a part of the three-loop $\beta$-function which depends on the Yukawa couplings has been found in \cite{Shakhmanov:2017soc,Kazantsev:2018nbl} by using the standard technique (see, e.g., \cite{Gates:1983nr,West:1990tg,Buchbinder:1998qv}) and in \cite{Stepanyantz:2019ihw} with the help of the method described in Sect. \ref{Section_Method}. The coincidence of the results confirms the correctness of this method. However, the above mentioned calculations were made in the minimal (Feynman) gauge. As a rule, the calculations in non-minimal gauges are much more complicated. Here we investigate the gauge dependence of the three-loop terms in the $\beta$-function which contain the Yukawa couplings.

\begin{figure}[h]
\begin{picture}(0,3)
\put(0.3,0.3){\includegraphics[scale=0.095,clip]{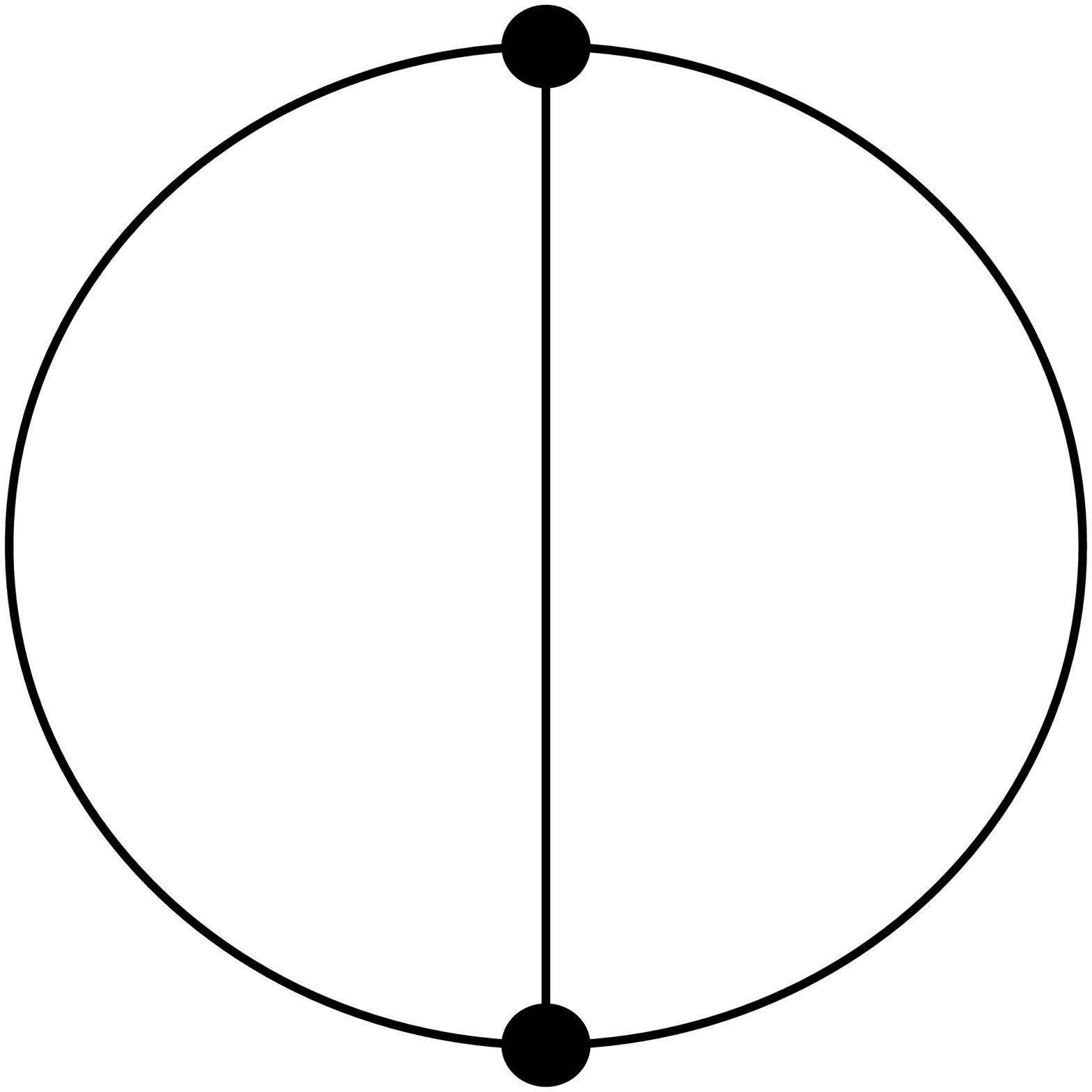}}
\put(0.0,2.0){(1)} \put(0.65,1.15){$K_\mu$} \put(1.55,1.15){$Q_\mu$}
\put(3.6,0.3){\includegraphics[scale=0.10,clip]{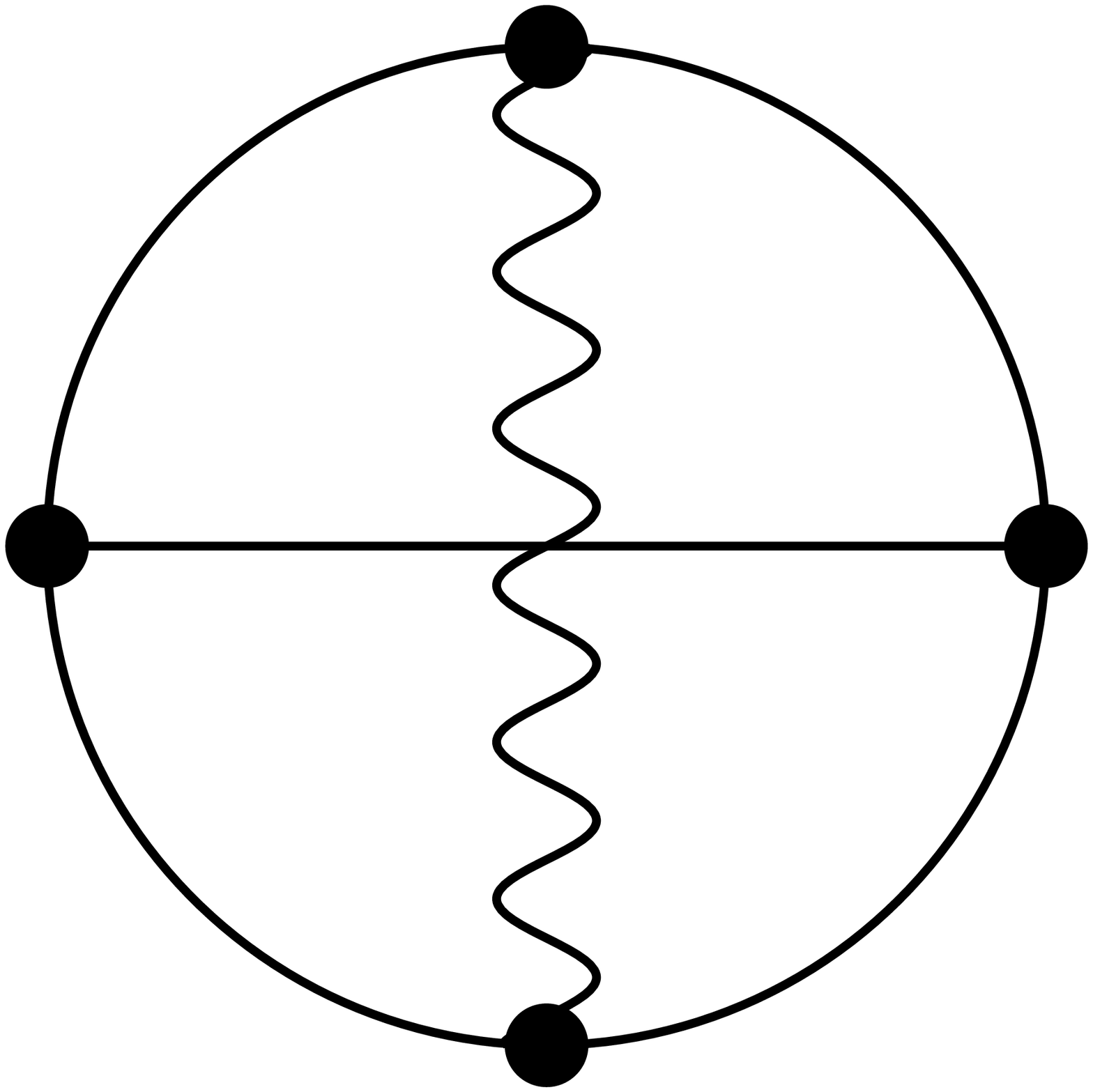}}
\put(3.2,2.0){(2)} \put(3.95,0.75){\small $K_\mu$}\put(3.95,1.35){\small $L_\mu$} \put(5.2,1.8){\small $Q_\mu$}
\put(7.0,0.3){\includegraphics[scale=0.10,clip]{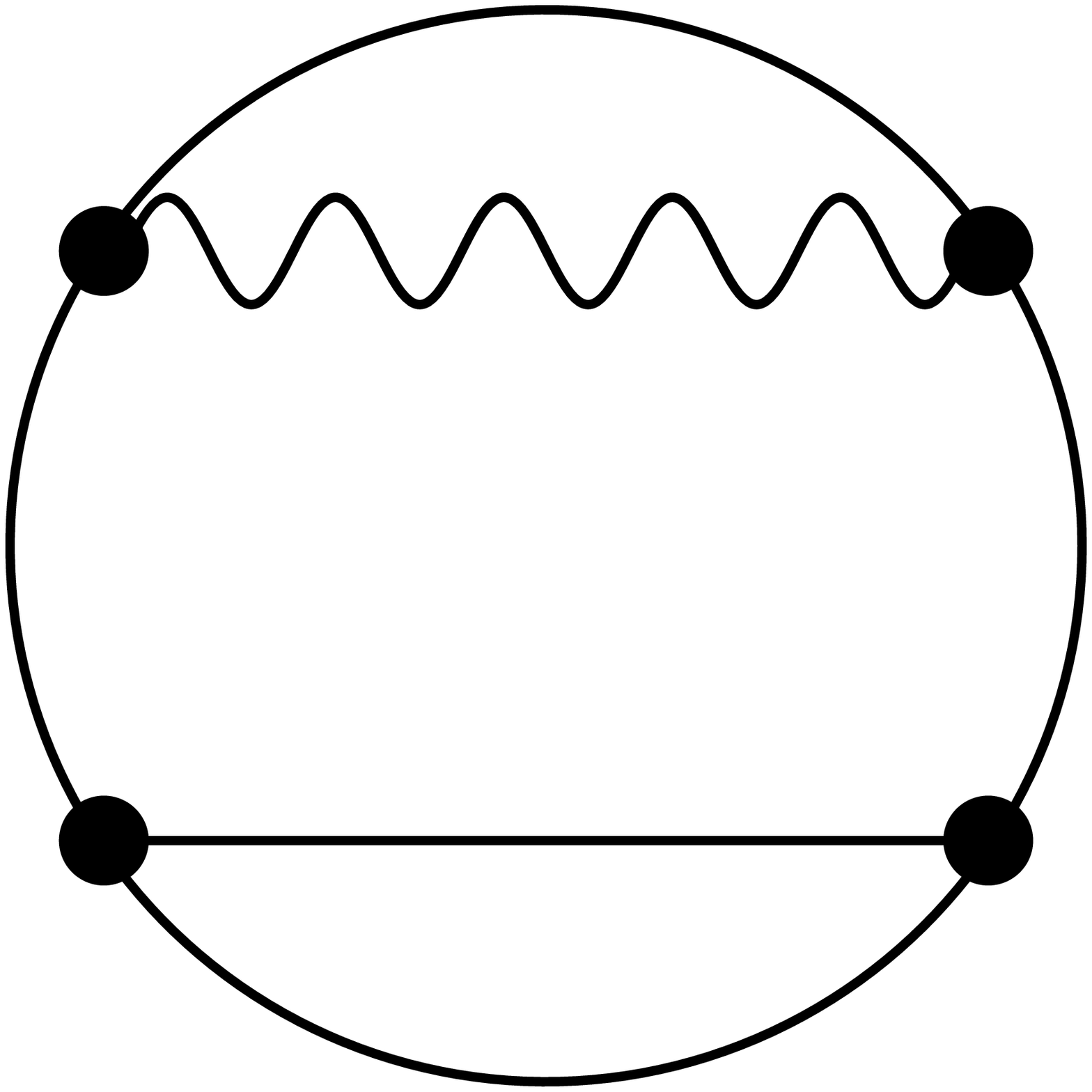}}
\put(6.6,2.0){(3)} \put(7.8,1.2){\small $K_\mu$}\put(7.8,-0.1){\small $L_\mu$} \put(8.8,1.05){\small $Q_\mu$}
\put(10.2,-0.1){\includegraphics[scale=0.21,clip]{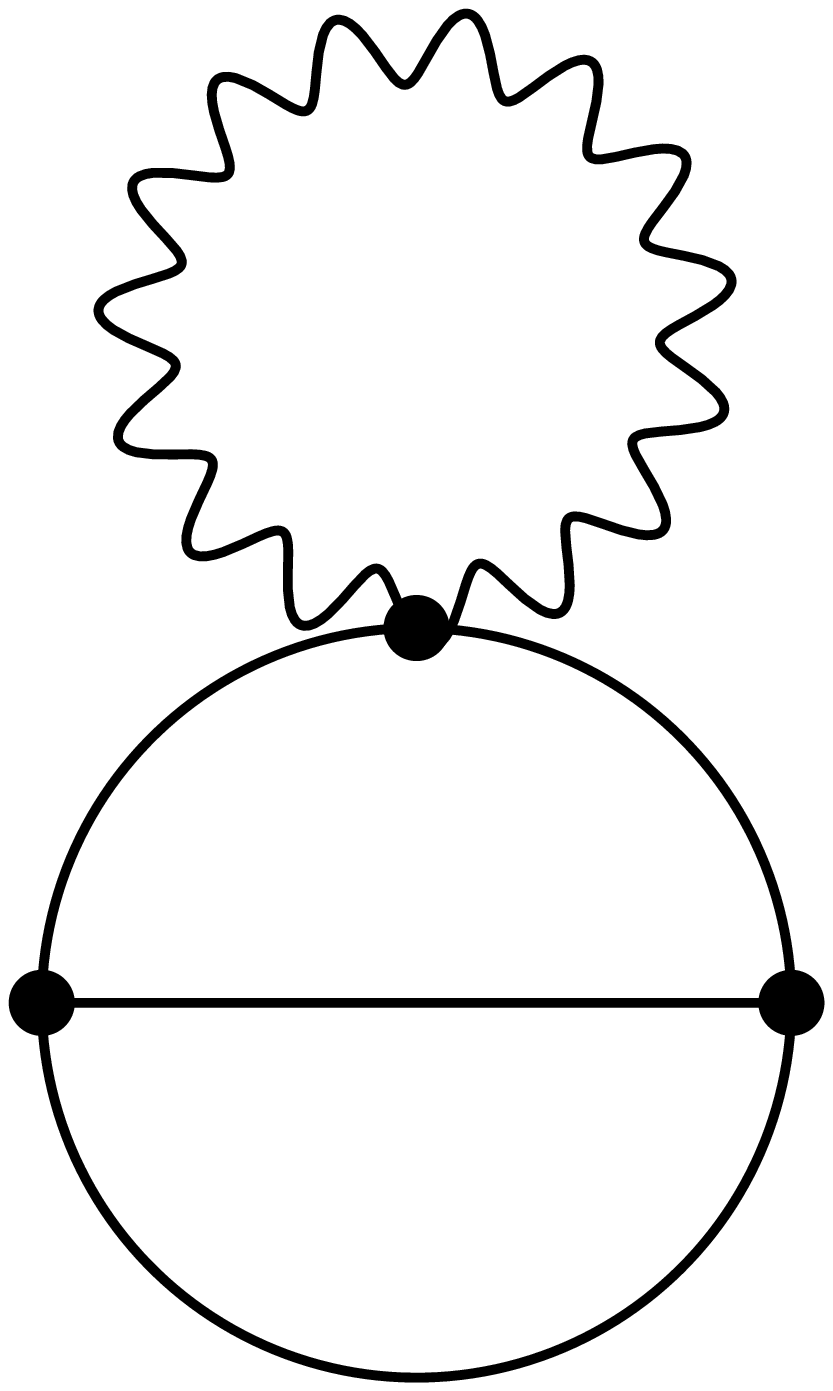}}
\put(9.7,2.0){(4)} \put(11.8,2.0){\small $K_\mu$}\put(11.78,0){\small $L_\mu$} \put(11.8,1.2){\small $Q_\mu$}
\put(13.4,0.3){\includegraphics[scale=0.10,clip]{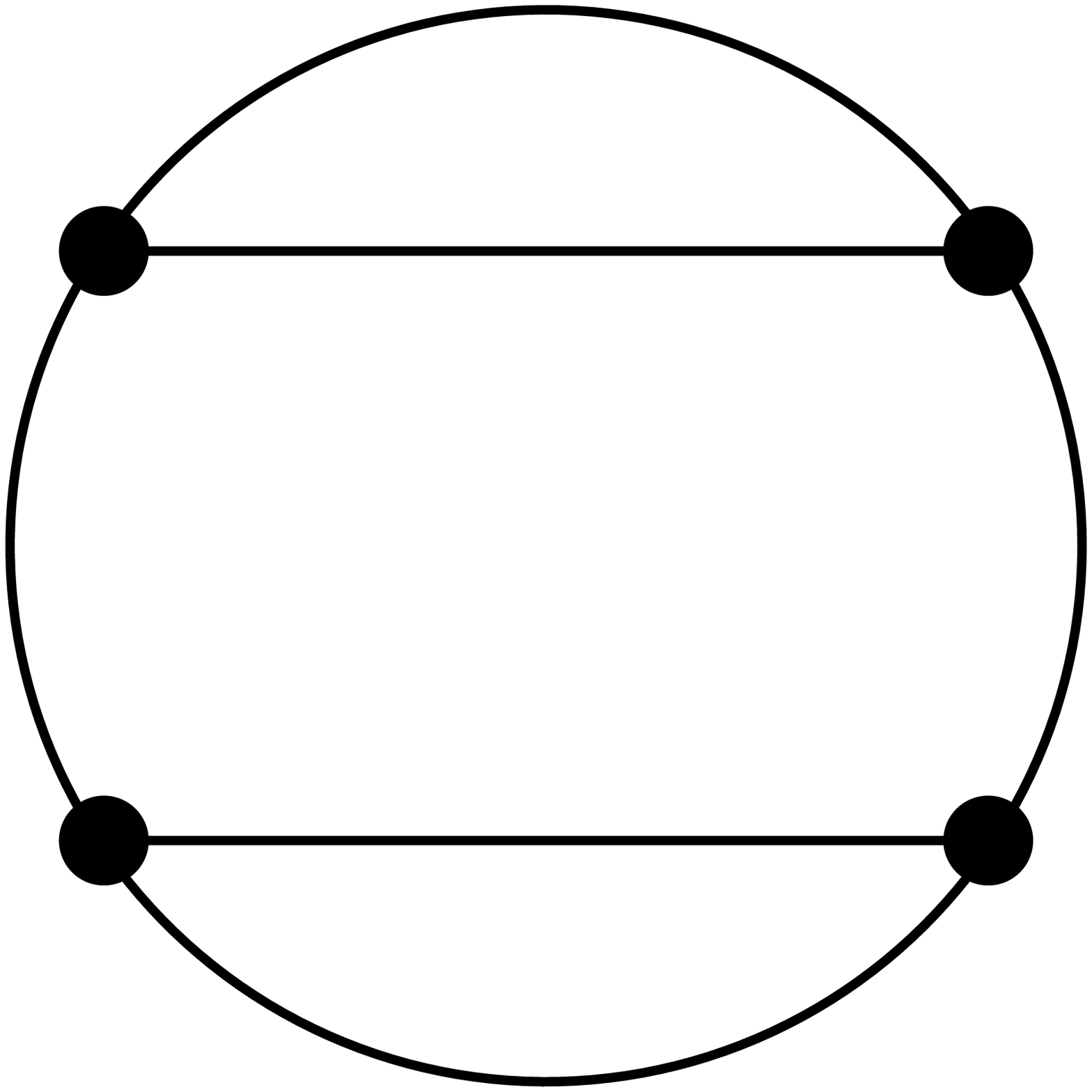}}
\put(13.0,2.0){(5)} \put(15.2,1.05){\small $K_\mu$}\put(14.1,-0.1){\small $L_\mu$} \put(14.1,2.2){\small $Q_\mu$}
\end{picture}
\caption{The vacuum supergraphs generating terms depending on the Yukawa couplings in the two-  and three-loop $\beta$-function.}\label{Figure_Yukawa_Supergraphs}
\end{figure}

Two- and three-loop contributions to the $\beta$-function which depend on the Yukawa couplings are generated by the supergraphs presented in Fig. \ref{Figure_Yukawa_Supergraphs}. As we have already mentioned, the usual superdiagrams are obtained from them by all possible attachments of two external $\bm{V}$-legs. The gauge dependence comes from the gauge propagator

\begin{equation}\label{Gauge_Propagator}
2i \bigg(\frac{1}{R\partial^2} - \frac{1}{16\partial^4} \Big(D^2 \bar D^2 + \bar D^2 D^2\Big)\Big(\frac{\xi_0}{K} - \frac{1}{R}\Big)\bigg)\delta^8_{xy}\delta^{AB},
\end{equation}

\noindent
where

\begin{equation}
\delta^8_{xy} = \delta^4(x^\mu-y^\mu)\delta^4(\theta_x-\theta_y).
\end{equation}

\noindent
The minimal gauge corresponds to the choice $K(x) = R(x)$ and $\xi_0=1$. In this case the expression (\ref{Gauge_Propagator}) takes a more simple form.

The gauge propagator is present only in the supergraphs (2) --- (4) in Fig. \ref{Figure_Yukawa_Supergraphs}, so that only their contributions to the function $\beta/\alpha_0^2$ can be gauge dependent. The results for them have been found using the method described in Sect. \ref{Section_Method} and are collected in Appendix \ref{Appendix_Supergraph_Results}. After some transformations the sum of the gauge dependent contributions can be presented in the form

\begin{eqnarray}\label{Gauge_Dependent_Terms}
&&\hspace*{-5mm} \sum\limits_{I=2,3,4} \frac{\Delta_I\beta}{\alpha_0^2} - \sum\limits_{I=2,3,4} \frac{\Delta_I\beta}{\alpha_0^2}\bigg|_{\xi_0=1;\,K=R}
= - \frac{8\pi}{r} \frac{d}{d\ln\Lambda} \int \frac{d^4K}{(2\pi)^4} \frac{d^4L}{(2\pi)^4} \frac{d^4Q}{(2\pi)^4} \frac{e_0^2}{K^4}  \Big(\frac{\xi_0}{K_K} - \frac{1}{R_K}\Big)\nonumber\\
&&\hspace*{-5mm} \times \Big(\lambda_0^{ijk} \lambda^*_{0ijl} \left(C(R)^2\right)_k{}^l - \lambda_0^{ipq} \lambda^*_{0imn} C(R)_p{}^m C(R)_q{}^n\Big) \Big(\frac{\partial^2}{\partial L^\mu \partial L_\mu} - 2 \frac{\partial^2}{\partial Q^\mu \partial L_\mu}\Big) \frac{1}{F_L L^2 F_{Q+K}}\nonumber\\
&&\hspace*{-5mm} \times \frac{1}{(Q+K)^2 F_{Q+L} (Q+L)^2} = 0,
\end{eqnarray}

\noindent
where we use the notations $r\equiv \mbox{dim}\, G$,\ \ $C(R)_i{}^j \equiv (T^A T^A)_i{}^j$,\ \ $F_K \equiv F(K^2/\Lambda^2)$, etc. To see that the result vanishes, we note that after the change of integration variables

\begin{equation}
L'_\mu \equiv L_\mu + Q_\mu;\qquad Q'_\mu \equiv - Q_\mu;\qquad K'_\mu \equiv - K_\mu
\end{equation}

\noindent
the differential operator present in the expression (\ref{Gauge_Dependent_Terms}) changes the sign,

\begin{equation}
\frac{\partial^2}{\partial L^\mu \partial L_\mu} - 2 \frac{\partial^2}{\partial Q^\mu \partial L_\mu} = - \frac{\partial^2}{\partial L'{}^\mu \partial L'_\mu} + 2 \frac{\partial^2}{\partial Q'{}^\mu \partial L'_\mu},
\end{equation}

\noindent
while the function of loop momenta remains the same. This implies that the expression (\ref{Gauge_Dependent_Terms}) is equal to 0, so that all gauge dependent contributions to the $\beta$-function (of the considered structure, in the considered approximation) cancel each other in agreement with the general theorems, see \cite{Batalin:2019wkb} and references therein. Therefore, the expression for the considered part of the $\beta$-function defined in terms of the bare couplings in the general $\xi$-gauge coincides with the one in the Feynman gauge. In the case of using the higher covariant derivative regularization the latter has been calculated in \cite{Kazantsev:2018nbl,Kazantsev:2020kfl} and is written as

\begin{eqnarray}\label{Beta_Resut}
&&\hspace*{-7mm}  \Delta\Big(\frac{\beta(\alpha_0,\lambda_0)}{\alpha_0^2}\Big) = - \frac{1}{8\pi^3 r} C(R)_j{}^i \lambda^*_{0imn} \lambda_0^{jmn} - \frac{\alpha_0 C_2}{16\pi^4 r} C(R)_j{}^i \lambda^*_{0imn} \lambda_0^{jmn} + \frac{\alpha_0}{16\pi^4 r} \left[C(R)^2\right]_j{}^i \nonumber\\
&&\hspace*{-7mm} \times \lambda^*_{0imn} \lambda_0^{jmn} \Big(1+A-B\Big) - \frac{\alpha_0}{8\pi^4 r} C(R)_j{}^i C(R)_l{}^n \lambda^*_{0imn} \lambda_0^{jml} \Big(1-A+B\Big) + \frac{1}{32\pi^5 r} C(R)_j{}^i \nonumber\\
&&\hspace*{-7mm} \times \lambda^*_{0iac} \lambda_0^{jab} \lambda^*_{0bde} \lambda_0^{cde}, \vphantom{\frac{1}{\pi^2}}
\end{eqnarray}

\noindent
where

\begin{equation}\label{Definition_Of_A_And_B}
A=\int\limits_0^\infty dx \ln x\, \frac{d}{dx}\frac{1}{R(x)};\qquad B = \int\limits_0^\infty dx \ln x\, \frac{d}{dx}\frac{1}{F^2(x)}.
\end{equation}

\noindent
The expression (\ref{Beta_Resut}) (as a part of the $\beta$-function defined in terms of the bare couplings) does not depend on a renormalization prescription which supplements the higher covariant derivative regularization and satisfies the NSVZ equation for all such prescriptions.

Starting from Eq. (\ref{Beta_Resut}) one can calculate the corresponding part of the $\beta$-function defined in terms of the renormalized couplings. The result can be found in \cite{Kazantsev:2018nbl,Kazantsev:2020kfl}. For a certain renormalization prescription it agrees with the expression obtained in the $\overline{\mbox{DR}}$ scheme \cite{Jack:1996vg}. In some other schemes (including the HD+MSL scheme \cite{Kataev:2013eta,Shakhmanov:2017wji,Stepanyantz:2017sqg}) it satisfies the NSVZ equation, see \cite{Kazantsev:2020kfl} for details.

\section{Conclusion}
\hspace*{\parindent}

In this paper for a general renormalizable ${\cal N}=1$ supersymmetric gauge theory with matter regularized by higher covariant derivatives and for an arbitrary $\xi$-gauge we have calculated a part of the three-loop $\beta$-function which depends on the Yukawa couplings. This has been done with the help of the method proposed in \cite{Stepanyantz:2019ihw,Kuzmichev:2019ywn}. Its main advantage is that instead of a very large number of supergraphs with two external lines of the background gauge superfield one should consider only relatively small number of vacuum supergraphs which should be modified in a special way. Then the algorithm described in Sect. \ref{Section_Method} produces a contribution to the $\beta$-function (defined in terms of the bare couplings) corresponding to the sum of all two-point superdiagrams generated by all possible attachments of two external gauge legs to the original vacuum supergraph. By construction, the result is written in the form of an integral of double total derivatives with respect to the loop momenta. Certainly, this algorithm drastically simplifies the technical part of the calculation. Moreover, it plays an important role in the all-loop perturbative derivation of the NSVZ equation, see \cite{Stepanyantz:2020uke}. However, it is so different from the standard technique that the various checks of it are highly desirable. The calculation made in this paper allowed to make a very nontrivial verification of this method. Really, in nonminimal gauges the calculations become much more complicated, but according to the general theorems the gauge dependence should disappear in the sum of all contributions. The results obtained in this paper by the new method demonstrate that the gauge dependence is present in the expressions for separate three-loop supergraphs, but vanishes in their sum. This fact can be considered as a highly nontrivial confirmation that this method works correctly  and can be used both for explicit calculations of quantum corrections and for proving general statements about their structure in supersymmetric theories.

\section*{Acknowledgments}
\hspace*{\parindent}

The work of K.S. was supported by Foundation for Advancement of Theoretical Physics and Mathematics ``BASIS'', grant  No. 19-1-1-45-1.

\appendix

\section{Contributions to the $\beta$-function corresponding to various vacuum supergraphs}
\hspace*{\parindent}\label{Appendix_Supergraph_Results}

Below we present the results for contributions to the $\beta$-function generated by the supergraphs (2) --- (4) in Fig. \ref{Figure_Yukawa_Supergraphs}. The expressions for the gauge independent supergraphs (1) and (5) coincide with the ones in the Feynman gauge and can be found, e.g., in \cite{Stepanyantz:2019ihw}.

\begin{eqnarray}
&& \frac{\Delta_2\beta}{\alpha_0^2} = - \frac{8\pi}{r} \frac{d}{d\ln\Lambda} \int \frac{d^4K}{(2\pi)^4} \frac{d^4L}{(2\pi)^4} \frac{d^4Q}{(2\pi)^4} e_0^2 \bigg\{\bigg[\frac{1}{2} \lambda_0^{ijk} \lambda^*_{0ijl} \left(C(R)^2\right)_k{}^l - \lambda_0^{ipq} \lambda^*_{0imn} C(R)_p{}^m \nonumber\\
&&\times C(R)_q{}^n\bigg] \frac{\partial}{\partial L^\mu}\Big(\frac{\partial}{\partial L_\mu}+\frac{\partial}{\partial Q_\mu}\Big) - \frac{1}{2} \lambda_0^{ijk} \lambda^*_{0ijl} \left(C(R)^2\right)_k{}^l \frac{\partial^2}{\partial Q^\mu \partial Q_\mu} - \frac{1}{2} C_2 \lambda_0^{ijk} \lambda^*_{0ijl} C(R)_k{}^l \nonumber\\
&& \times \frac{\partial}{\partial K^\mu} \Big(\frac{\partial}{\partial K_\mu} - \frac{\partial}{\partial Q_\mu}\Big) \bigg\} \bigg\{\frac{N(Q,K,L)}{K^2 R_K L^2 F_L Q^2 F_Q (Q+K)^2 F_{Q+K} (Q-L)^2 F_{Q-L} (Q+K-L)^2}\nonumber\\
&& \times \frac{1}{F_{Q+K-L}} + \Big(\frac{\xi_0}{K_K} - \frac{1}{R_K}\Big)\frac{2}{K^4 L^2 F_L (Q+K)^2 F_{Q+K} (Q-L)^2 F_{Q-L}}\bigg\};\\
&& \vphantom{\Big(}\nonumber\\
&& \frac{\Delta_3\beta}{\alpha_0^2} = - \frac{8\pi}{r} \frac{d}{d\ln\Lambda} \int \frac{d^4K}{(2\pi)^4} \frac{d^4L}{(2\pi)^4} \frac{d^4Q}{(2\pi)^4} e_0^2 \bigg\{\lambda_0^{ijk} \lambda^*_{0ijl} C_2 C(R)_k{}^l \frac{\partial}{\partial K^\mu} \Big(\frac{\partial}{\partial K_\mu} - \frac{\partial}{\partial Q_\mu}\Big)  + \lambda_0^{ijk} \nonumber\\
&& \times \lambda^*_{0ijl} \left(C(R)^2\right)_k{}^l \frac{\partial}{\partial Q^\mu} \Big(\frac{\partial}{\partial Q_\mu} - \frac{\partial}{\partial L_\mu}\Big) + \lambda_0^{ipq} \lambda^*_{0imn} C(R)_p{}^m C(R)_q{}^n \frac{\partial^2}{\partial L^\mu \partial L_\mu} \bigg\} \bigg\{\frac{1}{L^2 F_L Q^2 F_Q^2} \nonumber\\
&& \times \frac{1}{(Q+K)^2 F_{Q+K} (Q+L)^2 F_{Q+L}} \bigg[\frac{L(Q,Q+K)}{K^2 R_K} + \Big(\frac{\xi_0}{K_K} - \frac{1}{R_K}\Big)\frac{1}{K^4}\Big(Q^2 F_Q^2 + (Q+K)^2 \nonumber\\
&& \times F_{Q+K}^2\Big)\bigg] \bigg\};\\
&& \vphantom{\Big(}\nonumber\\
&& \frac{\Delta_4\beta}{\alpha_0^2} = \frac{8\pi}{r} \frac{d}{d\ln\Lambda} \int \frac{d^4K}{(2\pi)^4} \frac{d^4L}{(2\pi)^4} \frac{d^4Q}{(2\pi)^4} e_0^2 \bigg\{\lambda_0^{ijk} \lambda^*_{0ijl} C_2 C(R)_k{}^l \frac{\partial}{\partial K^\mu} \Big(\frac{\partial}{\partial K_\mu} - \frac{\partial}{\partial Q_\mu}\Big)  + \lambda_0^{ijk} \nonumber\\
&& \times \lambda^*_{0ijl} \left(C(R)^2\right)_k{}^l \frac{\partial}{\partial Q^\mu} \Big(\frac{\partial}{\partial Q_\mu} - \frac{\partial}{\partial L_\mu}\Big) + \lambda_0^{ipq} \lambda^*_{0imn} C(R)_p{}^m C(R)_q{}^n \frac{\partial^2}{\partial L^\mu \partial L_\mu} \bigg\} \bigg\{\frac{1}{L^2 F_L Q^2 F_Q^2} \nonumber\\
&& \times \frac{1}{(Q+L)^2 F_{Q+L}} \bigg[\frac{K(Q,K)}{K^2 R_K} + \Big(\frac{\xi_0}{K_K} - \frac{1}{R_K}\Big)\frac{F_{Q+K}}{K^4}\bigg] \bigg\}.
\end{eqnarray}

\noindent
In these expressions we use the notations (first introduced in \cite{Kazantsev:2018nbl})

\begin{eqnarray}
&& N(Q,K,L) \equiv L^2 F_{Q+K} F_{Q+K-L} - Q^2 \Big((Q+K)^2 - L^2\Big) F_{Q+K-L} \frac{F_{Q+K}-F_Q}{(Q+K)^2-Q^2}\nonumber\\
&& - (Q-L)^2 \Big((Q+K-L)^2-L^2\Big) F_{Q+K} \frac{F_{Q+K-L}-F_{Q-L}}{(Q+K-L)^2 - (Q-L)^2} + Q^2 (Q-L)^2 \qquad\nonumber\\
&&\times \Big(L^2 - (Q+K)^2 - (Q+K-L)^2\Big) \frac{F_{Q+K}-F_Q}{(Q+K)^2-Q^2} \frac{F_{Q+K-L}-F_{Q-L}}{(Q+K-L)^2-(Q-L)^2};\\
&& \vphantom{1}\nonumber\\
&& L(Q,Q+K)  \equiv F_Q F_{Q+K} + \frac{F_{Q+K} - F_Q}{(Q+K)^2 - Q^2} \Big(Q^2 F_Q + (Q+K)^2 F_{Q+K}\Big) + 2Q^2 \nonumber\\
&&\times (Q+K)^2 \Big(\frac{F_{Q+K}- F_Q}{(Q+K)^2 - Q^2}\Big)^2;\\
&& \vphantom{1}\nonumber\\
&& K(Q,K) \equiv \frac{F_{Q+K} - F_Q - 2Q^2 F_Q'/\Lambda^2}{(Q+K)^2 - Q^2} + \frac{2Q^2(F_{Q+K} - F_Q)}{\left((Q+K)^2 - Q^2\right)^2},
\end{eqnarray}

\noindent
where the prime denotes the derivative with respect to the argument $Q^2/\Lambda^2$.

\end{document}